\documentclass{aa} 
\usepackage{epsf} 
 
\newcommand{\gcc}{\mbox{~g\,cm$^{-3}$}} 
\newcommand{\beq}{\begin{equation}} 
\newcommand{\eeq}{\end{equation}} 
\newcommand{\req}[1]{Eq.~(\ref{#1})} 
\newcommand{\kB}{k_{\rm B}} 
\newcommand{\mel}{m_{\rm e}} 
\begin{document} 
 
  \thesaurus{12.          % A&A Section 12: Physical processes 
              (08.14.1;   % Stars: neutron 
               08.23.1;   % white dwarfs 
               02.04.1;   % Dense matter 
               02.03.2)   % Conduction 
             } 
 
% ************************************************************************* 
%                                 TITLE 
% ************************************************************************* 
\title{Transport properties of degenerate electrons in 
       neutron star envelopes and white dwarf cores} 
 
\author{A.Y.\ Potekhin\inst{1}\thanks{%
E-mail: palex@astro.ioffe.rssi.ru}, 
D.A.\ Baiko\inst{1}, 
P.\ Haensel\inst{2}, 
D.G.\ Yakovlev\inst{1}
}
\institute{
         Ioffe Physical-Technical Institute,
         Politekhnicheskaya 26, 194021 St.-Petersburg, Russia 
\and
	 N.\ Copernicus Astronomical Center, Bartycka 18,
	 PL 00-716 Warsaw, Poland}
\date{Received 3 December 1998 / Accepted 2 March 1999}
\offprints{A.Y.~Potekhin}
\titlerunning{Transport properties of degenerate electrons}
\authorrunning{A.Y.~Potekhin et al.}
\maketitle
 
% ************************************************************************* 
%                                ABSTRACT 
% ************************************************************************* 
\begin{abstract} 
New calculations of the
thermal and electrical electron conductivities
are performed for a broad range of physical parameters
typical for envelopes of neutron stars and cores of white dwarfs. 
We consider stellar matter composed of
astrophysically
important chemical elements from H to Fe in the density
range from $10^2$--$10^4\gcc$ up to $10^7$--$10^{10}\gcc$,
where atoms are fully ionized and electrons
are strongly degenerate.
We have used modified ion structure factors
suggested recently by Baiko et al.\ (1998).
In the ion liquid, these modifications
take into account, in an approximate way,
instantaneous electron-band structures
that reduce the electron-ion scattering rate.
In crystallized matter, the new
structure factors
include multi-phonon processes important
at temperatures not very much lower than the melting temperature
$T_{\rm m}$.
The transport coefficients obtained differ significantly
from those derived earlier
in the important temperature range
$T_{\rm m}/5 \la T \la 5 T_{\rm m}$.
The results of our numerical calculations
are fitted by analytical expressions
convenient for astrophysical applications.
\keywords{stars: neutron -- white dwarfs -- dense matter -- conduction} 
 
\end{abstract} 
% ************************************************************************* 
%                               TEXT BODY 
% ************************************************************************* 
%=Section 1=========================================================== 
\section{Introduction} 
\label{sect-intro} 
Thermal and electrical conduction in the envelopes of neutron stars
and the cores of white dwarfs 
plays crucial role in many aspects of evolution of these stars. 
Thermal conductivity 
is the basic quantity needed for 
calculating the relationship between the 
internal temperature of a neutron star 
and its effective surface temperature; this relationship 
affects thermal evolution of the neutron star and 
its radiation spectra 
(e.g., Gudmundsson et al.\ \cite{gpe83}; Page \cite{page97};
Potekhin et al.\ \cite{pcy97}, hereafter Paper~I). 
Electrical conductivity is the basic quantity 
used in calculations of magnetic-field evolution 
in neutron star crusts 
(e.g., Muslimov \& Page \cite{mp96}; 
Urpin \& Konenkov \cite{uk97}; Konar \& Bhattacharya \cite{kb97}). 
Thermal conductivity of degenerate  
matter is also an essential ingredient  
for white-dwarf pulsation modelling
(e.g., Fontaine \& Brassard \cite{fb94}). 
 
For applications one should know the
transport properties of dense stellar matter 
where electrons are strongly degenerate 
and form nearly ideal Fermi-gas, 
whereas ions are partially or fully ionized 
and form either a strongly coupled Coulomb liquid 
or a Coulomb crystal. 
Under such conditions, electrons are usually most important 
charge and heat carriers, 
and the electrical and thermal conductivities
are mainly determined by
electron scattering off ions (hereafter, \emph{ei} scattering). 
 
The conductivities of degenerate electrons due to \emph{ei} scattering 
were studied in a number of papers. The general formalism, 
based on a variational method (Ziman \cite{Ziman}), 
has been developed by Flowers \& Itoh (\cite{fi76}) 
(see references to earlier results therein). 
Their numerical results, however, have been critically revised by 
Yakovlev \& Urpin (\cite{yu80}), 
who developed a simple 
analytical description of the conduction 
due to \emph{ei} scattering in dense Coulomb plasmas. 
The results by Yakovlev \& Urpin (\cite{yu80}) 
were confirmed in detailed calculations 
for both solid (Raikh \& Yakovlev \cite{ry82}) 
and liquid (Itoh et al.\ \cite{itoh-ea83};  
Nandkumar \& Pethick \cite{np84}) regimes. 
Later Yakovlev (\cite{yak87}) calculated the electron transport 
coefficients in the liquid phase 
taking into account non-Born corrections. 
In Paper~I, the authors performed extensive calculations 
of the same coefficients including additionally the effects of 
responsive electron background on the ion structure factors. 
 
Itoh et al.\ (\cite{itoh-ea84,itoh-ea93}) improved 
the results by Yakovlev \& Urpin (\cite{yu80})  
and Raikh \& Yakovlev (\cite{ry82}) 
in the solid regime 
by including the nuclear form factor $F(q)$ 
(where $\hbar q$ is a momentum transfer), 
in order to take into account finite sizes of atomic nuclei, 
and studied the role of 
the Debye--Waller factor $\mathrm{e}^{-2W(q)}$ 
(e.g., Kittel \cite{Kittel}), which describes 
reduction of \emph{ei} scattering rate in a crystal 
due to lattice vibrations. 
The Debye--Waller factor proved to be important 
at temperatures close to the 
melting temperature of a Coulomb crystal, 
and at sufficiently high densities, 
where zero-point vibrations are large. 
Later Baiko \& Yakovlev (\cite{bya95,bya96}) made detailed 
Monte Carlo and analytical calculations of the electron 
transport in crystalline dense matter 
including the nuclear form factor and the Debye--Waller factor; 
they fitted the results by simple analytical expressions. These 
results were in reasonable agreement 
with those obtained by  
Itoh et al.\ (\cite{itoh-ea84,itoh-ea93}). 
 
Nevertheless the transport theory developed in the cited 
articles possessed one 
important drawback: it predicted unrealistically large 
jumps (by a factor of 2--4) of the transport coefficients 
at the melting point in spite of the fact that 
many other physical 
properties of ion liquid and solid were 
% there 
very similar. 
This indicated that the theory was incomplete 
and had to be improved. 
 
The improvements have been suggested recently by
Baiko et al.\ (\cite{bkpy}) (hereafter Paper~II). 
In the solid regime, 
multi-phonon processes have been included
in the electron-phonon scattering, whereas all previous 
calculations used the one-phonon approximation. 
Furthermore, it has been noted in Paper~II, that the   
static structure factor of ions
conventionally used in the liquid regime
may require modification when applied to calculation 
of the electron scattering processes 
because of appearance of incipient ordered structures.
The authors suggested an approximate treatment of this effect  
by subtracting a certain part from the 
static structure factor. Both modifications affect significantly 
the electron transport properties near the melting point 
and reduce the jumps of the transport coefficients. 
 
In this paper, we apply the results by Baiko et al.\ (\cite{bkpy})
to calculation of the electron electrical and thermal conductivities
in a wide range of physical parameters 
typical for the envelope of a neutron star or 
the core of a white dwarf.
We also derive an 
effective \emph{ei} scattering potential that 
yields 
the conductivities in a simple analytical form, 
reproducing the numerical results 
within, at most, a few tens of percent 
in the whole of the physically meaningful range of 
the plasma parameters. 
Finally, we discuss the 
main features of the electron transport properties, 
and the role of various electron scattering mechanisms. 
 
The paper is organized as follows. 
In Sect.~\ref{sect-basics} we introduce basic definitions  
and give a brief overview of the main features 
of electron conduction in different regimes. 
In Sect.~\ref{sect-outer} we describe our calculation 
of the electron transport coefficients 
and propose a fit to these coefficients. 
Numerical results are discussed 
in Sect.~\ref{sect-discussion}. 
 
%=Section  =========================================================== 
\section{Dense degenerate matter} 
\label{sect-basics} 
\subsection{Equilibrium properties} 
\label{sect-struct} 
Consider fully ionized degenerate stellar matter in the density range 
from about $10^2$--$10^4\gcc$ 
to $\sim10^{10}\gcc$. 
For simplicity, we assume that there is one ion species 
at any given $\rho$ and $T$. 
 
The state of degenerate electrons can be described by 
the Fermi momentum $p_{\rm F}$ or wave number $k_{\rm F}$: 
\beq 
    p_{\rm F}=\hbar k_{\rm F} 
    =\hbar\, (3 \pi^2 n_{\rm e})^{1/3} = \mel c x_{\rm r}, 
\label{xr} 
\eeq 
where $n_{\rm e}$ is the electron number density, 
$\mel$ is the electron mass,  
$x_{\rm r} \approx 1.009 \, (\rho_6 Z/A)^{1/3}$ 
is the relativistic parameter, 
$Z$ is the ion charge number, 
$A$ is the atomic weight, 
% is the total number of baryons per ion, 
and $\rho_6$ is mass density $\rho$ 
in units of $10^6\gcc$. The electron degeneracy 
temperature is  
\beq 
     T_{\rm F} = (\epsilon_{\rm F} - \mel c^2)/\kB  
        \approx 5.93 \times 10^9 
      \left(\sqrt{1+x_{\rm r}^2}-1 \right)\mbox{~K}, 
\eeq 
where $\kB$ is the Boltzmann constant 
and $\epsilon_{\rm F}=\mel c^2\, \sqrt{1+x_{\rm r}^2} 
\equiv \mel^* c^2$  
is the electron Fermi energy.  
Our analysis 
is thus limited by the condition $T \ll T_{\rm F}$. 
It is also restricted to 
temperatures $T \la 5 \times 10^9$~K at which atomic 
nuclei do not dissociate. 
Electrostatic screening properties of the electron gas are
characterized by the Thomas--Fermi wave number $k_{\rm TF}$:
\beq
     k^2_{\rm TF}=4\pi e^2 \, {\partial n_{\rm e}\over\partial \mu}
\approx
     {\alpha_{\rm f} \over\pi}\,{\sqrt{1+x_{\rm r}^2} \over x_{\rm r}}
           \,(2 k_{\rm F})^2 ,
\eeq
where $\mu\approx\epsilon_{\rm F}$ is the electron chemical potential
and $\alpha_{\rm f} = e^2/\hbar c = 1/137.036$ 
is the fine-structure constant.
 
The state of ions (atomic nuclei) can be conveniently 
specified by the Coulomb plasma parameter 
\beq 
    \Gamma = {(Ze)^2\over \kB  T a_{\rm i}} 
  \approx {22.75\, Z^2 \over T_6}\left({\rho_6\over A}\right)^{1/3}, 
\label{Gamma} 
\eeq 
where $e$ is the elementary charge, 
$a_{\rm i}=[3/(4\pi n_{\rm i})]^{1/3}$ is the ion-sphere radius, 
$n_{\rm i}=n_{\rm e}/Z$ is the number density of ions, 
and $T_6$ is temperature 
in units of $10^6$~K. If $\Gamma \ll 1$, ions are weakly coupled 
and form the Boltzmann gas. For $\Gamma \ga 1$, they constitute 
a strongly coupled liquid. 
Freezing occurs 
at a temperature $T_{\rm m}$ which corresponds to 
$\Gamma = \Gamma_{\rm m}$. 
For classical ions, 
$\Gamma_{\rm m}=172$, whereas quantum effects 
(zero-point ion vibrations) suppress the freezing and increase 
$\Gamma_{\rm m}$ (Nagara et al.\ \cite{nnn87}). 
The quantum effects become important at $T\ll T_{\rm p}$, where 
\beq 
      T_{\rm p}  =  \hbar \omega_{\rm p} / \kB  
          \approx 7.832 \times 10^6 \, (Z/A)\rho_6^{1/2}~{\rm K} 
\label{T_p} 
\eeq 
is the ion plasma temperature, 
$\omega_{\rm p}  = ( 4 \pi Z^2 e^2 n_{\rm i} / m_{\rm i} )^{1/2}$ 
is the ion plasma frequency, 
% $A$ is the atomic weight of an ion, 
$m_{\rm i}=A m_{\rm u}$ is the ion mass, and 
$m_{\rm u}=1.6605\times10^{-24}$~g is the atomic mass unit. 
Under realistic conditions, 
the quantum effects strongly suppress crystallization of hydrogen 
and helium plasmas, but do not affect significantly the melting 
of carbon and heavier elements (e.g., Chabrier \cite{chabr93}). 
However, they affect the properties of matter of any 
composition at $T\ll T_{\rm p}$. 
 
% Subsection ==================================================== 
\subsection{Transport coefficients and structure factors} 
\label{sect-sfact} 
Electrical ($\sigma$) and thermal ($\kappa$) conductivities 
of degenerate electrons 
can be conveniently expressed through  
effective electron collision frequencies, 
$\nu_\sigma$ and $\nu_\kappa$, as (e.g., Ziman \cite{Ziman}; 
Yakovlev \& Urpin \cite{yu80}) 
\begin{equation} 
   \sigma={n_{\rm e} e^2 \over \mel^* \nu_\sigma}, 
\qquad 
   \kappa = {\pi^2  
       \kB^2 T n_{\rm e} 
       \over 3 \mel^* \nu_\kappa}. 
\label{s,k} 
\end{equation} 
The collision frequencies are reduced to sums of 
partial collision frequencies associated with relevant
electron scattering mechanisms which can be studied 
separately. 
This approximation is accurate to $\sim1\%$
in the case of strongly degenerate electrons 
(e.g., Ziman \cite{Ziman}; Lampe \cite{lampe68}).
 
In the solid phase 
at $T \ll T_{\rm p}$, 
where the frequency 
of \emph{ei} collisions $\nu_{\sigma,\kappa}^{ei}$ is strongly reduced,
the electron transport is limited by scattering 
off various irregularities of the crystalline structure. 
The particular case of ion impurities,
occasionally embedded in the lattice,
was studied in detail by Itoh \& Kohyama (\cite{ik93})
and will be taken into account in this article;
in this case
charge and heat transport are determined by a single
collisional frequency $\nu_{\rm imp}$.
On the other hand, in the liquid phase,
at not very strong electron degeneracy for 
$Z \la 6$ the thermal (but not the electrical)
conductivity may be affected by electron-electron
(\emph{ee}) collisions.  The \emph{ee} collision frequency
$\nu^{ee}$
was evaluated, e.g.,   
by Urpin \& Yakovlev (\cite{uy80}), Timmes (\cite{timm92}), 
and in Paper~I [see \req{ee} below].
These results for electron-electron and electron-impurity scattering 
are not modified by the present consideration. 
The total effective collision frequencies are 
 $\nu_\sigma = \nu_\sigma^{ei}$,
$\nu_\kappa=\nu_\kappa^{ei}+\nu_\kappa^{ee}$ in the liquid
and $\nu_{\sigma,\kappa} = \nu_{\sigma,\kappa}^{ei}+\nu_{\rm imp}$
in the solid.

We will focus on \emph{ei} scattering.  
In a weakly coupled ion gas, $\Gamma \ll 1$,
collective effects lead to a dynamical ion screening of \emph{ei}
interaction which can be described by
the dynamical dielectric function formalism
(e.g., Williams \& DeWitt \cite{wdw69}).
In a strongly coupled ion liquid, it is customary
to use the static structure factor of ions
in order to describe the correlation effects (Hubbard \cite{hubb66}).
This description, however, does not apply to quantum liquids,
such as H or He at high densities (e.g., Chabrier \cite{chabr93}).
In crystalline matter, \emph{ei} interaction
is adequately described in terms of absorption and emission of
phonons (Abrikosov \cite{abrik61}). 
The description can be realized using
a dynamical structure factor of ions (Flowers \& Itoh \cite{fi76}).

The \emph{ei} collision frequencies 
can be expressed 
through dimensionless 
{\it Coulomb logarithms} $\Lambda_{\sigma,\kappa}$ 
(cf.\ Yakovlev \& Urpin \cite{yu80}): 
\begin{equation} 
   \nu^{ei}_{\sigma,\kappa} = 
         { 4 \pi Z^2 e^4 n_{\rm i} \over p_{\rm F}^2 v_{\rm F} } 
          \, \Lambda_{\sigma,\kappa}
         = { 4 Z \epsilon_{\rm F} 
         \over 3 \pi \hbar }  
     \,\alpha_{\rm f}^2 \Lambda_{\sigma,\kappa}, 
\label{nusk} 
\end{equation} 
where $v_{\rm F}=p_{\rm F}/m_{\rm e}^\ast$ 
is the electron Fermi velocity. 
 
For a strongly coupled plasma of ions ($\Gamma \ga 1$), 
the Coulomb logarithms 
calculated in the variational approach
(with the simplest trial functions, Ziman \cite{Ziman})
in the Born approximation read 
\begin{equation} 
    \Lambda_{\sigma,\kappa} = \int_{q_0}^{2k_{\rm F}} 
    {\rm d}q\,q^3 u^2(q) \, S_{\sigma,\kappa}(q) \, 
    \left[1 - {v_{\rm F}^2 \over c^2} 
    \left({q\over2k_{\rm F}}\right)^2  \right], 
\label{L} 
\end{equation} 
where $q_0$ is  
the cutoff parameter, equal to zero for the liquid phase  
and to the equivalent radius of the Brillouin zone 
$q_{\rm B}  = (6 \pi^2 n_i)^{1/3}$ in the solid phase.
The latter cutoff filters out 
\emph{umklapp} electron-phonon processes,
which operate at $q>q_{\rm B}$ and give the main contribution 
to $\Lambda_{\sigma,\kappa}$, from
normal processes, that take place at $q<q_{\rm B}$ and
are negligible under the conditions of study
(e.g., Baiko \& Yakovlev \cite{bya95}).
Furthermore, 
$u(q)\equiv |U(q)|/(4\pi Z e^2)$, 
$U(q)$ is the Fourier transform of the 
elementary \emph{ei} scattering potential, 
the factor in square brackets describes kinematic suppression 
of backward scattering of relativistic electrons  
(e.g., Berestetskii et al.\ \cite{LaLi-QED}), 
and $S_{\sigma,\kappa}(q)$ are the  
{\it effective} static structure factors which take into account 
ion correlations, as discussed below.  
 
In order to take into account corrections to the Born approximation,
we multiply additionally the integrand in \req{L}
by the ratio of the exact cross section of Coulomb electron 
scattering to the Born cross section.
This approximate treatment of non-Born corrections
was proposed by Yakovlev (\cite{yak87}) and used 
in Paper~I. 
The corrections are significant 
for $Z \ga 20$ and $\rho \ga 10^6\gcc$. 
 
In the case of Coulomb scattering, one has 
\beq 
    u(q)= {F(q) \over q^2 |\varepsilon(q)|}, 
\label{u} 
\eeq 
where $\varepsilon(q)$ is a static longitudinal 
dielectric function of the electron gas 
(Jancovici \cite{janco62}), describing electron screening of 
the scattering potential. 
The static ($\omega \to 0$) approximation of electron screening 
is adequate as long as typical energies transferred 
$\hbar\omega$ are small compared with $\hbar q v_{\rm F}$, 
which is the case, since the momentum
transfer $\hbar q \sim p_{\rm F}$. 
At densities $\rho < 10^{10}$ g cm$^{-3}$ and under the condition of 
full ionization, one can safely set the ion form factor $F(q)$
equal to 1, which corresponds to point-like nuclei.  

The structure factors in \req{L} are given by (Paper~II) 
\begin{eqnarray} 
 S_\sigma(q)\,\, &=\,\,& 
    \int_{-\infty}^{+\infty} {\rm d}\omega \, S''(q,\omega) \, 
    {z \over 1- \mathrm{e}^{-z} }, 
\label{Ssk-sigma} 
\\
   S_\kappa(q) \,\, &=\,\,& S_{\sigma}(q) + 
          \left( {3 k_{\rm F}^2\over q^2}-\frac12\right) \delta S_\kappa(q), 
\\ 
   \delta S_\kappa(q) &=&   
    \int_{-\infty}^{+\infty} {\rm d}\omega \, S''(q,\omega) \, 
       { z^3 \over 1-\mathrm{e}^{-z} }. 
\label{Ssk-delta} 
\end{eqnarray} 
In this case $z=\hbar \omega / (\kB T)$
and $S''(q,\omega)$ is the inelastic part 
of the total dynamical structure factor 
$S(q,\omega) = S''(q,\omega)+S'(q) \delta(\omega)$,
whereas the elastic part, $S'(q)\delta(\omega)$,
describes Bragg diffraction.  
When interaction of electrons  
with a crystalline lattice is considered, 
the Bragg diffraction leads to appearance 
of electron band structure (Bloch states) but does 
not contribute to electron transport 
(e.g., Flowers \& Itoh \cite{fi76}). 
If $z$ values ``allowed'' by $S''(q,\omega)$ 
in Eqs.~(\ref{Ssk-sigma})--(\ref{Ssk-delta}) 
are small, as it happens for scattering in a classical Coulomb system 
[i.e., at $T \ga T_{\rm p}$; cf.\ Eq.\ (\ref{SqOas})], we can 
pull the other functions containing $z$ out of the integral 
setting $z=0$. 
Then $\delta S_\kappa(q)$ vanishes, and 
$S_\sigma(q)=S_\kappa(q)=S''(q)$, 
where 
$S''(q)=\int_{-\infty}^{+\infty}S''(q,\omega){\rm\,d}\omega$ 
is the inelastic part of the static structure factor. 
In this case the variational solution that is used in \req{L} 
becomes exact.
 
In the liquid phase, 
$\Gamma < \Gamma_{\rm m}$, it is only 
the static structure factor 
in the classical regime 
$S(q) = \int^{+\infty}_{-\infty} {\rm d}\omega S(q,\omega)$ 
(e.g., Young et al.\ \cite{ycdw91} and references therein) 
that has been determined quite accurately. 
Thus we will consider only \emph{classical} liquids, $T \ga T_{\rm p}$.  
In the solid  
phase $S(q,\omega)$ was calculated with reasonable accuracy 
in Paper~II, which enables us to study the 
transport properties of quantum and 
classical solids. 
 
%=Section 3=========================================================== 
\section{Calculation of electron conductivities} 
\label{sect-outer} 
% 
% ===================================================== 
\subsection{Solid phase} 
\label{sect-solid} 
The dynamical structure factor 
of a Coulomb crystal has been calculated
in Paper~II
in the harmonic-lattice approximation, 
taking into account explicitly multi-phonon processes. 
For $q>q_{\rm B}$ the inelastic part may be written as 
\begin{eqnarray} 
S''(q,\omega)\,\, &\!\!\! =& \!\!\!{\mathrm{e}^{ - 2W(q)- 
   \hbar \omega/(2\kB T)} \over 2 \pi} 
   \int^{+\infty}_{-\infty} \!\!\!\!\!\! {\rm d}t \,  
      \mathrm{e}^{-{\rm i} \omega t } K(q,T,t), 
\label{S''}
\\ 
  K(q,T,t) &=& 
   \exp \left[ {\hbar q^2 \over 2 m_{\rm i}} 
   \left\langle {\cos{\omega_\nu t} \over \omega_\nu \sinh{(z_\nu/2)}} 
   \right\rangle_{\rm ph} \right] - 1  , 
\label{K} 
\end{eqnarray} 
where $\langle \ldots \rangle_{\rm ph}$ 
denotes averaging over the phonon spectrum in the first 
Brillouin zone, and 
\begin{equation} 
    W(q) = { 3\hbar\over2 m_{\rm i} } \left\langle 
      {(\vec{q}\cdot\vec{e}_\nu)^2 \over\omega_\nu} 
         \left(\bar{n}_\nu+\frac12\right)   
       \right\rangle_{\rm ph} . 
\label{DW} 
\end{equation} 
In this case, 
$\nu \equiv (\vec{Q},s)$, 
$s=1,2,3$ enumerates phonon polarizations, 
\vec{Q} is a phonon wave vector, 
$\vec{e}_\nu$ the polarization vector, $\omega_\nu$ 
the frequency, and 
$\bar{n}_\nu = \left( \mathrm{e}^{z_\nu}-1 \right)^{-1}$ is 
the mean number of phonons, $z_\nu=\hbar \omega_\nu/(\kB T)$. 
For the lattice types of interest [e.g., body centred cubic (bcc) 
or face centred cubic (fcc) ones], 
$W(q) = r_T^2 q^2 / 6$, 
where $r_T^2$ 
is the mean-squared ion displacement. 
Thus $W(q)$ does not depend on the orientation of $\vec{q}$. 
An analytical fit to $W(q)$ was 
proposed by Baiko \& Yakovlev (\cite{bya95}): 
\beq 
     W(q)= \frac14 \alpha_1  
     \left( u_{-1} \, \mathrm{e}^{-9.1 \eta} + 2 \eta \, u_{-2} \right), 
\label{DWfit} 
\eeq 
where $\eta = T/T_{\rm p}$, 
$u_n= \langle (\omega_\nu / \omega_{p})^n \rangle_{\rm ph}$ 
is a frequency moment of Coulomb lattice ($u_{-2}=13.0$ and 
$u_{-1}=2.8$ for bcc lattice, cf.\ Pollock \& Hansen \cite{ph73}), 
and 
\beq 
   \alpha_1 = \alpha_0 \, {q^2\over 4 k_{\rm F}^2}, 
\qquad 
  \alpha_0 = {4 k^2_{\rm F} a^2_i \over 3 \Gamma \eta} = 
        1.683 \sqrt{ {x_{\rm r} \over  A Z} }. 
\label{alpha0} 
\eeq 

It is possible to derive an asymptote of $S''(q,\omega)$,
\req{S''},
for $T \gg T_{\rm p}$ (classical solid) 
and for $|\omega|\gg\omega_{\rm p}$.
Expanding $\cos{\omega_\nu t} \approx 
1 - (\omega_\nu t)^2/2$
and $\sinh{(z_\nu/2)} \approx z_\nu/2$ and noting that
the second term in \req{K} does not contribute into $S''(q,\omega)$
at these frequencies, we obtain 
\begin{equation} 
   S''(q,\omega) \approx  {1 \over \sqrt{\pi} \, q v_T} 
         \exp \left( - {\omega^2 \over q^2 v^2_T} 
           -{\hbar \omega \over 2\kB T} \right), 
\label{SqOas} 
\end{equation} 
where $v_T = \sqrt{2 \kB T/m_{\rm i}}$ is the thermal ion velocity. 
Thus in a classical solid the collisional energy transfer
is limited either by $\hbar q v_T$ 
or by $\hbar\omega_{\rm p}$, in both cases being smaller than
$\kB T$. Therefore only the values $z\ll1$ contribute to the integrals
(\ref{Ssk-sigma}), (\ref{Ssk-delta}) in this case.
 
In the general case of classical or quantum Coulomb crystals,
the effective structure factors (\ref{Ssk-sigma}) and (\ref{Ssk-delta})  
can be written as (Paper~II) 
\begin{eqnarray} 
   S_{\sigma}(q)\,\, &\! =& {1  \over 2 }\, \mathrm{e}^{-2W(q)} 
    \int_{-\infty}^{+\infty} { {\rm d}x \over \cosh^2 x } 
     \, K(q,T,t), 
\label{Ss-sol} 
\\ 
   \delta S_\kappa(q) &=& 
   \mathrm{e}^{-2W(q)} 
         \int_{-\infty}^{+\infty} {\rm d}x \, 
        {1-2 \sinh^2 x \over \cosh^4 x}  \,K(q,T,t), 
\label{Ssk-sol} 
\end{eqnarray} 
where $x=\pi t T/\hbar$. We have calculated $S_\sigma(q)$ 
and $\delta S_\kappa(q)$ for the bcc lattice 
and fitted them by the expressions 
\begin{eqnarray} 
 S_\sigma(q)\,\, &\! =& \mathrm{e}^{-2W(q)} \left( \mathrm{e}^{2W_1(q)} - 1 \right), 
\label{Sfit} \\ 
 \delta S_\kappa(q) &=& \alpha_1 \left[ 
     { 91 \, \eta^2 \, \mathrm{e}^{-2W(q)} \over (1+ 111.4 \eta^2)^2} \right. 
\nonumber \\ 
 & + & \left. {0.101 \, \eta^4 \over (0.06408+\eta^2)(0.001377+\eta^2)^{3/2}} 
      \right], 
\label{deltaSfit} \\ 
   W_1(q) &=& {\alpha_1 \, u_{-2} \, \eta^2 \over 2 \, 
           \sqrt{\eta^2 + (u_{-2} / 117)^2} }. 
\nonumber 
\end{eqnarray} 
These fits cover a wide range of the parameters, 
$0.001 \le \eta \le 10$ 
and $0\leq\alpha_1 \leq 0.3$, sufficient 
for calculation of transport coefficients. 
Equations (\ref{Sfit}) and (\ref{deltaSfit}) 
reproduce also the asymptotes of the effective 
structure factors at low and high $\eta$ which can be 
obtained from Eqs.\ (\ref{Ss-sol}) and (\ref{Ssk-sol}).  
The maximum fit error, equal to 
4\%, occurs for $\alpha_1=0.001$, and $\eta=0.04$. 
 
We have also calculated the effective structure factors for 
fcc Coulomb lattice and the results appear to be almost 
indistinguishable from those obtained for bcc lattice. 
Therefore, the electron transport coefficients are 
insensitive to the lattice type,\footnote{%
The results by Baiko \& Yakovlev (1995, 1996)
for the fcc lattice, which led to a different conclusion,
were inaccurate due to an error in a Brillouin zone
integration scheme for this lattice.
}
and we will calculate them for bcc lattice. 
 
%%%%%%%%%%%%%%%%%%%%%%%%%%%%%%%%%%%%%%%%%%%%%%%%%%%%%%%%%%%%%%% 
\subsection{Liquid phase} 
\label{sect-liquid} 
In the classical liquid, we employ the static structure factor, 
$S(q)$,  
obtained by Rogers and DeWitt (unpublished) 
for the one-component classical plasma of ions in a rigid 
electron background by solving the modified hypernetted-chain 
(MHNC) equations (Rosenfeld \& Ashcroft \cite{ra79}), 
and fitted by Young et al.\ (\cite{ycdw91}) in the range 
$1 \leq \Gamma \leq 225$. 
For calculating \emph{ei} scattering rates, 
we have modified these structure factors  
by subtracting the contribution of elastic scattering 
as prescribed in Paper~II: 
\begin{equation} 
    S_{\sigma,\kappa}(q) = S(q) - 
    \mathrm{e}^{-2W(q)} (2\pi)^3 n_i \sum_{\vec{G}\neq 0} 
    \overline{\delta(\vec{q}-\vec{G})}, 
\label{Ssk-liq} 
\end{equation} 
where the sum is over all non-zero  
reciprocal lattice vectors $\vec{G}$,  
and the upper bar means averaging over orientations of \vec{q}. 
This modification is meant to 
account for instantaneous electron band-structures which 
emerge in a strongly coupled Coulomb liquid because of 
local temporary crystal-like ordering. 
In this context, the choice of the lattice type and corresponding 
vectors $\vec{G}$ is 
ambiguous, but the Coulomb logarithms 
are insensitive to it. 

Since our consideration is based on the classical static
structure factor $S(q)$,
we obtain $\Lambda_\sigma = \Lambda_\kappa$ 
and $\nu^{ei}_\sigma=\nu^{ei}_\kappa$ in the liquid regime. 
This is 
justified because we typically have $T \ga T_{\rm p}$, 
for the cases under study (see below). Nevertheless we could easily 
incorporate the quantum effects into the calculation, 
were the quantum dynamical structure factors $S(q,\omega)$ 
available for ion fluid. 

Note that in Paper~I
the Coulomb logarithms in the ion liquid were calculated with  
$S(q)$ obtained for a polarizable electron background 
including the local field corrections. 
The effect of the response of the background appeared to be noticeable 
for H and He plasmas only.  
We neglect this effect in the present 
calculations because our consideration of ion liquid
for light elements is approximate anyway
due to the neglect of quantum corrections to $S(q)$.
 
For light elements, the highest temperatures 
corresponding to the liquid regime, $\Gamma \geq 1$, 
are much below $T_{\rm F}$. Accordingly, 
there exists a temperature range  
where $T<T_{\rm F}$ and $\Gamma < 1$. 
In this interval  
the formalism of the effective structure factors 
does not provide an accurate treatment of ion screening. 
For $\Gamma \leq 0.25$, the Coulomb logarithms 
were calculated in Paper~I taking 
into account dynamical character of ion screening 
at $\Gamma \ll 1$ and $k_{\rm TF}\ll k_{\rm F}$
(e.g., Williams \& DeWitt \cite{wdw69}):
\begin{eqnarray}
   \Lambda_{\sigma,\kappa} &=& 
     \ln  \left( {2k_{\rm F} \over q_{\rm D}} \right) -
       {\zeta \over 2} \, \ln   \left( {1 \over \zeta } \right) +
    { 1+ \zeta \over 2 } \, \ln  \left( {1 \over 1+ \zeta} \right)
\nonumber \\&&
       -  v_{\rm F}^2 / (2 c^2) + \beta/2 ,
\label{ei-DeWitt}
\end{eqnarray} 
$q_{\rm D} = \sqrt{3\Gamma}/a_i$ is the inverse Debye 
screening length of ion gas,\footnote{%
Definitions of the screening parameters 
below Eq.~(15) of Paper~I,
corresponding to the present \req{ei-DeWitt},
contained several misprints.
Correct definitions are reproduced here.
}
$\zeta=(k_{\rm TF}/q_{\rm D})^2$,
and $\beta = \pi\alpha_{\rm f} Z v_{\rm F}/c$;
$\beta/2$ is a lowest-order non-Born correction in the 
weak electron-screening approximation (Yakovlev \cite{yak87}).

The 
Coulomb logarithms in the transition domain from 
weak ($\Gamma \ll 1$) to strong 
($\Gamma \ga 1$) ion coupling can be calculated using 
the formalism by Boerker et al.\ (\cite{brdw82}). We do not 
apply this formalism, but the Coulomb logarithms 
calculated at $\Gamma \leq 0.25$ and at $\Gamma \geq 1$ 
converge nicely and can be fitted in a unified manner. 
This convergence deteriorates at lower $\rho \sim 10$--100 
$\gcc$, because electron screening ceases to be weak 
($k_{\rm TF}\sim k_{\rm F}$), 
and \req{ei-DeWitt} becomes inaccurate. 
 
%%%%%%%%%%%%%%%%%%%%%%%%%%%%%%%%%%%%%%%%%%%%%%%%%%%%%%%%%%%%%%% 
\subsection{Numerical results and fitting formulae} 
\label{sect-numres} 
Using the effective structure factors $S_{\sigma,\kappa}(q)$ 
described above, 
we have calculated 
the Coulomb logarithms $\Lambda_{\sigma, \kappa}$ for $Z$ from 
1 to 26 and for mass numbers $A$ 
corresponding to the most abundant isotopes.  
The mass density $\rho$ varied 
from $10\gcc$ for $Z=1,2$ and from $100\gcc$ for $Z \ge 3$ 
to $10^{10}\gcc$; the coupling parameter $\Gamma$ varied from 1 to  
$10^4$ for $Z<20$ and to $10^5$ for $20\leq Z\leq26$. 
The physically meaningful domain of the parameters 
is constrained by several conditions. First, it is assumed 
that the atoms are fully ionized (for a treatment of the case of 
partial ionization, see below). Secondly, light elements 
are not present at very high densities and temperatures
since they burn into heavier ones.  
Thirdly, our calculation in the \emph{liquid} phase 
is confined to 
the classical regime ($T\ga T_{\rm p}$), where 
the quantum corrections to the ion structure 
factors are neglected. 
Fourthly, electrons are assumed to be degenerate ($T \ll T_{\rm F}$). 
Finally, the present formalism appears to be invalid at 
very low temperatures, $T \ll T_{\rm p} Z^{1/3} e^2/(\hbar v_{\rm F})$, 
where the electron band-structure effects strongly reduce 
\emph{ei} scattering rate (e.g., Raikh \& Yakovlev \cite{ry82}), 
which is not taken into account in Eq.\ (\ref{L}). 
 
For practical applications,  
especially for modelling thermal and magnetic  
evolution of the neutron stars, 
as well as pulsations of the white dwarfs, it is desirable to have  
analytical formulae for the transport coefficients, 
rather than tables, graphs or cumbersome theoretical expressions. 
Analytical formulae for $\sigma$ and $\kappa$ 
were presented in several papers
based on the earlier theoretical results
described in Sect.~\ref{sect-intro}
(Flowers \& Itoh \cite{fi81}; 
Yakovlev \& Urpin \cite{yu80}; Itoh et al.\ \cite{itoh-ea93}; 
Baiko \& Yakovlev \cite{bya95}; Paper~I). 
As our present results  
are significantly different,
we propose new analytical expressions, 
which combine reasonable accuracy with simplicity. 
 
Instead of constructing ad hoc 
fits to the numerical values of 
$\sigma$ and $\kappa$, 
we have chosen to devise an \emph{effective 
ei-scattering potential} that would 
allow us to perform explicit analytical integration in \req{L} 
and that would reproduce correctly the familiar limiting cases: 
the case of Debye ion screening in a weakly-coupled 
plasma and the case of scattering by high-temperature phonons. 
In the first case,  
$u^2(q) S(q)$ in \req{L} should be replaced by $(q^2+q_{\rm s}^2)^{-2}$,  
where $q_{\rm s}$ is the inverse screening length. 
In the second case (at $T_{\rm p} \la T \leq T_{\rm m}$), 
\emph{ei} scattering rate is determined by 
$u^2(q) S''(q)$, where 
$u^2(q) \approx (q^2+k_{\rm TF}^2)^{-2}$, 
and $S''(q)$ 
is the approximate effective static structure factor 
(Paper~II) which can be written as 
$S''(q) \approx 1- \exp[- u_{-2} a_{\rm i}^2 q^2 /(3 \Gamma)]$. 
This approach ensures that the analytical limits 
mentioned above are reproduced 
automatically by the fit expression. 
 
We propose the following form of the effective  
potential in \req{L}: 
\beq 
   \left[ u^2 (q) S_{\sigma,\kappa}(q) \right]_{\rm eff} = 
         {  1- \mathrm{e}^{-w(q)} 
        \over (q^2 + q_{\rm s}^2)^2 } 
       \, G_{\sigma,\kappa}(\eta,\beta) D(\eta), 
\label{Ufit} 
\eeq 
where $\mathrm{e}^{-w(q)}$ plays role of an effective Debye--Waller factor
at large $\Gamma$ and is negligible at $\Gamma < 1$,
$q_{\rm s}$ is an effective screening wave number, 
$D(\eta)$ is associated with the quantum correction 
to the Debye--Waller factor,
and $G_{\sigma,\kappa}(\eta,\beta)$  
is a phenomenological factor that describes 
reduction of thermal ion displacements in quantum solid 
at $T\la T_{\rm p}$
and contains non-Born corrections 
expressed through the argument $\beta$ [see \req{ei-DeWitt}].
Our numerical results 
are reproduced with the following choice of $q_{\rm s}$,
$w(q)$ and $G_{\sigma,\kappa}(\eta,\beta)$: 
\begin{eqnarray}
     q_{\rm s}^2 &=& (q_{\rm i}^2 + k_{\rm TF}^2)\,\mathrm{e}^{-\beta},
\label{qs2}
\\
     q_{\rm i}^2(\Gamma) &=&  
     q_{\rm D}^2\,(1+0.06\,\Gamma) \, \mathrm{e}^{-\sqrt\Gamma}, 
\label{qi} 
\\
    w(q) &=& u_{-2} (q / q_{\rm D})^2 \,(1+\beta/3),
\\
    G_\sigma(\eta,\beta) &=& 
        {\eta \over \sqrt{\eta^2 + \eta_0^2}} \,(1+0.122\beta^2), 
\quad 
    \eta_0 = {0.19 \over Z^{1/6}}, 
\\ 
     G_\kappa(\eta,\beta) &=& G_\sigma(\eta,\beta) 
       + 0.0105 \left( 1-Z^{-1} \right) 
         \nonumber\\ &&\times 
%	\left[ 1+(v_{\rm F}/c)^3 \beta \right]
%       \eta \, (\eta^2 + 0.0081)^{-3/2}, 
	\left[ 1+\left({v_{\rm F}\over c}\right)^3 \beta \right]
      { \eta \over (\eta^2 + 0.0081)^{3/2}}, 
\\ 
         D(\eta) &=& \exp  \left( - \alpha_0 u_{-1} \, 
                         \mathrm{e}^{-9.1 \eta}/4 \right), 
\label{Deta} 
\end{eqnarray} 
where $\alpha_0$ is given by Eq.\ (\ref{alpha0}). 
Inserting \req{Ufit} into \req{L}, 
we arrive at the analytical expressions for the Coulomb logarithms 
presented in the Appendix.

The typical error of our fits in the physically reasonable 
range of parameters is 3\% (maximum 6\%)
for $Z\geq20$ and gradually increases
with decreasing $Z$. The maximum error occurs for low $Z$
at the melting point in the high-density region,
where our formulae interpolate across the 
conductivity jump discussed in the next section.
For instance, a typical error for carbon plasma is 
8\% and the maximum is 22\% at 
the highest density ($\rho=10^{9}\gcc$) and $\Gamma=\Gamma_{\rm m}$. 

Finally, let us outline the case of multicomponent 
ion mixtures. Actually the case deserves a separate study 
and we treat it approximately here. 
At least for $T \ga T_{\rm p}$, 
it would be a good approximation 
to replace $Z^2 n_i \Lambda_{ei} \to 
\sum_j Z_j^2 n_j \Lambda_{ej}$ in Eq.\ (\ref{nusk}), 
where summation is over all ion species $j$, and 
the Coulomb logarithm $\Lambda_{ej}$ depends generally on $j$. 
In a strongly coupled ion system we 
recommend 
to calculate $\Lambda_{ej}$ from Eqs.\ (\ref{Lambdafit}), 
(\ref{qs2})--(\ref{Deta}) 
with $Z=Z_j$ and 
$\Gamma_j = Z_j^{5/3}e^2 
(4 \pi n_{\rm e}/3)^{1/3}(\kB T)^{-1}$ 
(the ion-coupling 
parameter for ion species $j$).
The latter expression is prompted by the ``additivity rule''
that is highly accurate for thermodynamic functions
of multicomponent ion mixtures (Hansen et al.\ \cite{hansen-ea77}).

Another option is 
to adopt the widely used
mean-ion approximation. In the latter approximation, 
the plasma is treated 
as a mixture of electrons and one ionic species, with an effective 
charge $eZ_{\rm eff}$ equal to an average charge  
of all ions at different ionization stages. 
The mean-ion approximation can be used also in the regime 
of partial ionization (cf.\ Paper~I).
 
% Fig.  %%%%%%%%%%%%%%%%%%%%%%%% FIGURE %%%%%%%%%%%%%%%%%%%%%%%%%%%%%% 
\begin{figure}[ht] 
\vspace{-0.4cm} 
\begin{center} 
 \leavevmode 
 \epsfysize=100mm 
\epsffile[30 8 380 425]{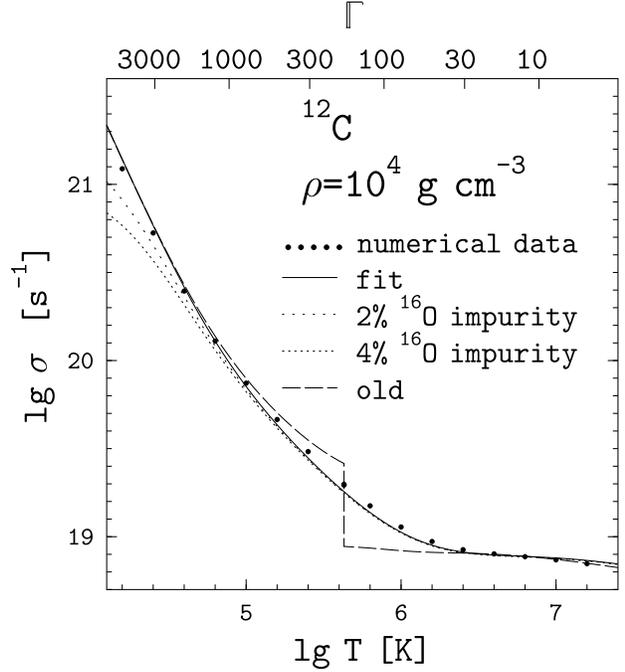}
\end{center} 
\vspace{-0.4cm} 
\caption[]{ 
Dependence of the electrical 
conductivity produced by \emph{ei} scattering 
on temperature (lower horizontal scale) or 
Coulomb coupling parameter (upper horizontal scale) 
in carbon plasma at $\rho=10^4$ g cm$^{-3}$. 
Filled circles show the present numerical results, 
and the solid line is given by our fitting formula. 
The dashed line is obtained under traditional assumptions 
(see the text) and exhibits a jump at the melting point. 
Dotted lines are obtained including electron scattering 
by $^{16}$O impurities with concentrations 
$x_{\rm imp}=0.02$ and 0.04. 
} 
\label{C4-s} 
\end{figure} 
% 
% 
%%%%%%%%%%%%%%%%%%%%%%%%%%%%%%%%%%%%%%%%%%%%%%%%%%%%%%%%%%%%%%% 
\section{Discussion of the results} 
\label{sect-discussion} 
% 
% Fig.  %%%%%%%%%%%%%%%%%%%%%%%% FIGURE %%%%%%%%%%%%%%%%%%%%%%%%%%%%%% 
\begin{figure}[ht] 
\vspace{-0.4cm} 
\begin{center} 
 \leavevmode 
 \epsfysize=100mm 
\epsffile[30 8 380 425]{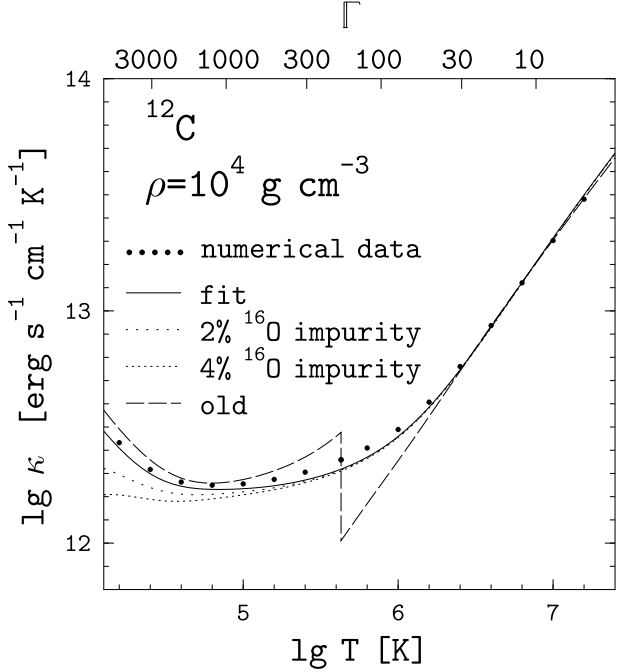}
\end{center} 
\vspace{-0.4cm} 
\caption[]{ 
Same as in Fig.\ \protect{\ref{C4-s} } 
but for the thermal conductivity. 
} 
\label{C4-k} 
\end{figure} 

Fig.~\ref{C4-s} shows the temperature dependence of 
the electrical conductivity 
of degenerate electrons in a carbon plasma at $\rho = 10^4$ g cm$^{-3}$. 
Fig.~\ref{C4-k} shows the same dependence of 
the thermal conductivity. Upper horizontal scale 
indicates corresponding values of the ion coupling 
parameter $\Gamma$. 
Since the ion charge number is rather low, $Z=6$, 
non-Born corrections are insignificant. All the data 
presented in the figures, except dotted lines, show the 
conductivities produced solely by \emph{ei} scattering. 
 
Filled circles display our present numerical values 
of the conductivities, while solid lines are the 
analytical fits. Dashed lines show the `old' conductivities 
calculated using the approximations which have been 
widely employed in the previous works (Sect.\ \ref{sect-intro}): 
the one-phonon approximation in the solid phase and 
use of the total (inelastic + elastic) ion structure 
factor in the liquid phase. One can see large jumps 
of the `old' curves at the melting point. These jumps 
were present for all elements and 
for all plasma parameters, and they were typically 
a factor of 2--4 in magnitude (Itoh et al.\ \cite{itoh-ea93}). 
The modification of the structure factors 
improves the treatment of the conductivities both 
in the liquid and solid phases of strongly-coupled ion system 
(Sects.\ \ref{sect-solid} and \ref{sect-liquid})  
and makes the jumps almost invisible 
for different chemical elements in a broad range of 
densities. 
This has allowed us to produce the 
unified fits which are equally 
valid in solid and liquid matter. 
 
% Fig.  %%%%%%%%%%%%%%%%%%%%%%%% FIGURE %%%%%%%%%%%%%%%%%%%%%%%%%%%%%% 
\begin{figure}[ht] 
\vspace{-0.4cm} 
\begin{center} 
 \leavevmode 
 \epsfysize=100mm 
\epsffile[30 8 380 425]{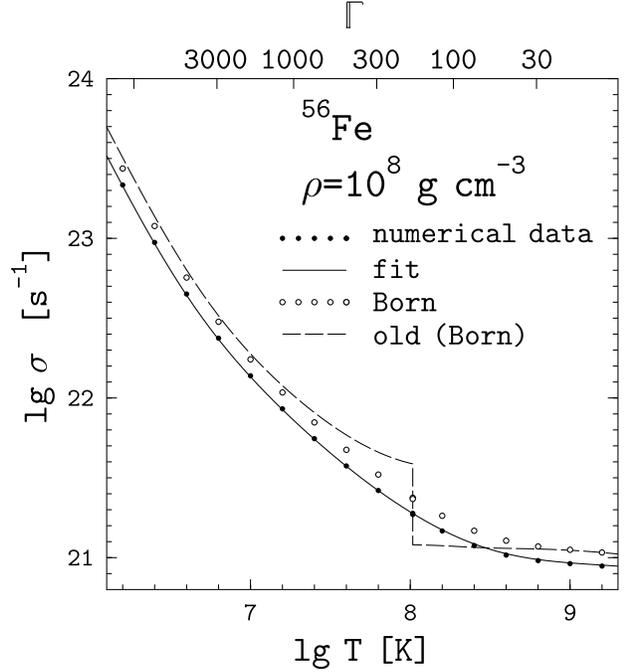} 
\end{center} 
\vspace{-0.4cm} 
\caption[]{ 
Electrical conductivity produced by \emph{ei} scattering 
in Fe matter at $\rho=10^8$ g cm$^{-3}$ vs 
temperature or Coulomb coupling parameter. 
Filled circles show the present numerical results, 
and the solid line is the fit (non-Born corrections included). 
Open circles are the present results but in the Born approximation. 
The dashed line is the old result 
in the Born approximation. 
} 
\label{Fe8-s} 
\end{figure} 

Nevertheless, our calculations do show large jumps 
of the conductivities at the melting point 
at high densities, where zero point 
vibrations become important. We suggest that 
these jumps are artificial and 
come from using classical ion structure 
factors in ion liquid (Sect.\ \ref{sect-liquid})  
under the conditions 
in which quantum effects in liquid are really important. 
On the other hand, the quantum effects are properly included 
in our calculations for crystalline matter. Since 
the numerical data used for constructing the fitting 
formulae include both phases, liquid and solid, 
the fitting of these data by the unified analytical 
expressions shifts the conductivities 
in the liquid phase closer to those 
in the solid phase. Therefore we expect that 
the fits in the high-density ion liquid 
give more reliable electron conductivities than our 
original numerical data. This assumption will be checked 
in the future when the structure factors 
in ion liquid are calculated taking into account the quantum effects. 
 
Dotted lines in Figs.\ \ref{C4-s} and \ref{C4-k} show 
the effect of another electron scattering mechanism -- 
scattering by charged impurities (Sect.\ \ref{sect-sfact}).  
We have assumed an admixture of oxygen nuclei with 
concentrations 2\% and 4\%. 
Electron-impurity scattering is seen to have little 
effect on the conductivities at high temperatures, 
but it becomes dominant scattering mechanism at 
$T \ll T_{\rm p}$. 
 
% 
% Fig.  %%%%%%%%%%%%%%%%%%%%%%%% FIGURE %%%%%%%%%%%%%%%%%%%%%%%%%%%%%% 
\begin{figure}[ht] 
\vspace{-0.4cm} 
\begin{center} 
 \leavevmode 
 \epsfysize=100mm 
\epsffile[30 8 380 425]{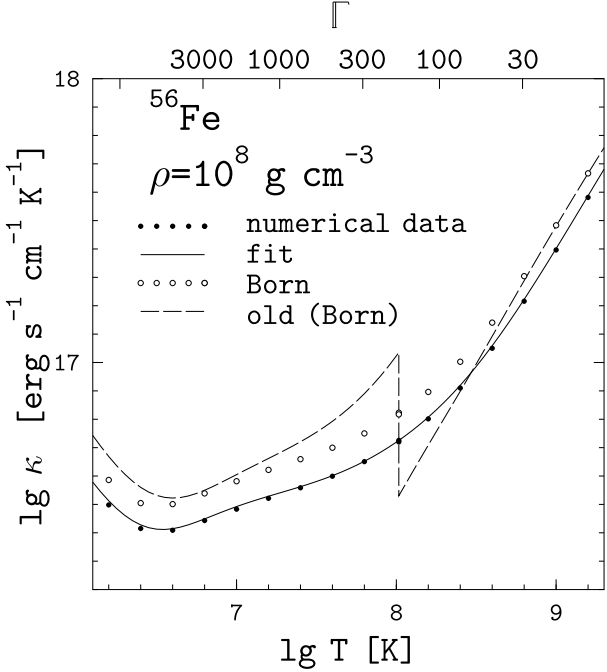} 
\end{center} 
\vspace{-0.4cm} 
\caption[]{ 
Same as in Fig.\ \protect{\ref{Fe8-s} } 
but for the thermal conductivity. 
} 
\label{Fe8-k} 
\end{figure} 

Figs.~\ref{Fe8-s} and \ref{Fe8-k} 
are analogous to Figs.\ \ref{C4-s} and \ref{C4-k} 
and show the temperature dependence 
of the conductivities produced by \emph{ei} scattering 
in iron plasma at $\rho = 10^8$ g cm$^{-3}$. 
Filled dots are our numerical results and 
solid lines are the fits.  
One can again see large jumps of 
the traditional conductivities at the melting point, 
and the smooth character of the improved curves. 
For elements  
with high $Z$ (like Fe), the non-Born corrections become 
important in a dense plasma (Yakovlev \cite{yak87}). 
To illustrate this effect, open circles display 
results of our calculations 
in the Born approximation. The non-Born corrections are 
seen to increase the \emph{ei} collision frequency 
(decrease the conductivities) by about 20--30\%. 
The corrections become lower when 
density decreases below $10^6$ g cm$^{-3}$. Dashed 
lines show the `old' conductivities calculated  
in the Born approximation. 
The divergence between the `old' and new
results in the Born approximation, seen in Fig.~\ref{Fe8-k}
at relatively low temperatures,
is caused by a not very accurate determination of
the low-temperature ($T\ll T_{\rm p}$) asymptote 
of an effective electron-phonon potential 
in the `old' results by Baiko \& Yakovlev (\cite{bya95}).

% Fig.  %%%%%%%%%%%%%%%%%%%%%%%% FIGURE %%%%%%%%%%%%%%%%%%%%%%%%%%%%%% 
\begin{figure}[ht] 
\begin{center} 
 \leavevmode 
 \epsfysize=95.5mm 
\epsffile[27 10 377 407]{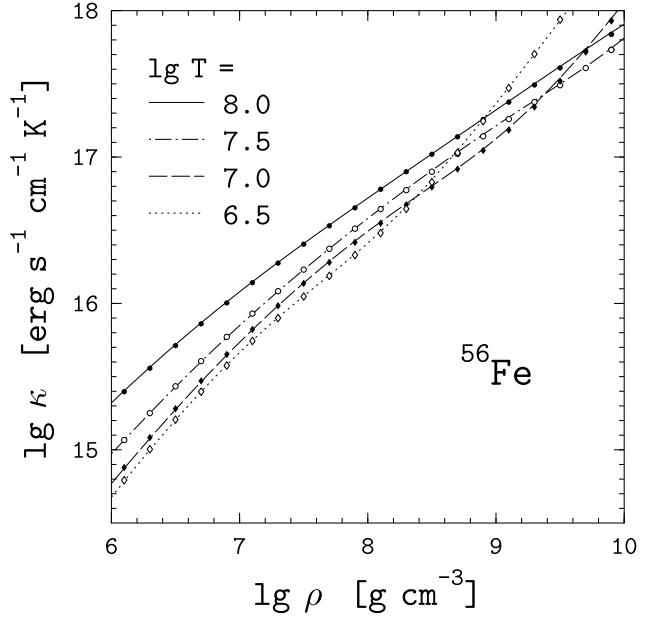}
\end{center} 
\vspace{-0.4cm} 
\caption[]{ 
Thermal conductivity vs density in iron plasma 
at several $T$. Various symbols show numerical results, and 
curves are the fits. Conductivity due to \emph{ee} collisions 
is included but is entirely unimportant. Impurities 
are neglected. 
} 
\label{Fe-r} 
\end{figure} 

\begin{figure}[ht] 
%\vspace{-0.4cm} 
\begin{center} 
 \leavevmode 
 \epsfysize=95.5mm 
\epsffile[27 10 377 407]{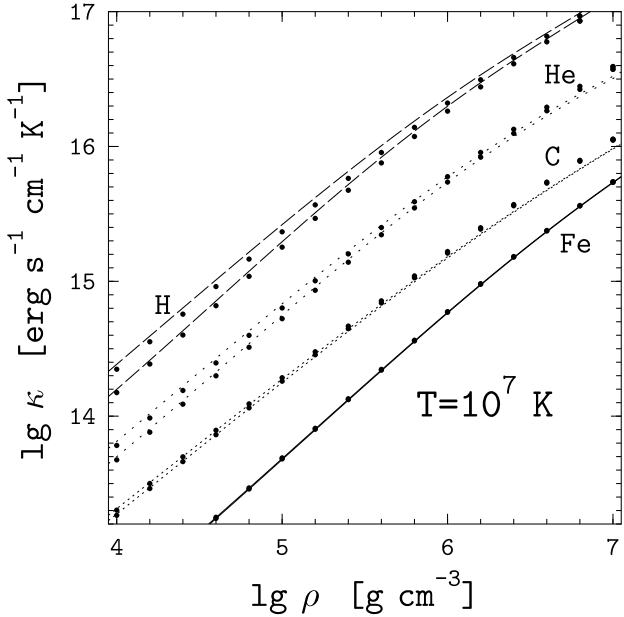}
\end{center} 
\vspace{-0.4cm} 
\caption[]{ 
Thermal conductivity vs density in matter of 
different chemical composition (H, He, C, Fe) 
at $T=10^7$ K. Filled circles are numerical values 
and curves are fits. Lower curves and circles 
are calculated including \emph{ei} and \emph{ee} collisions, while 
upper curves and circles take account of \emph{ei} collisions alone. 
} 
\label{all-r} 
\end{figure}

The density dependence of the thermal 
conductivity at several values of $T$ 
is displayed in Figs.\ \ref{Fe-r} and \ref{all-r}. 
Fig.~\ref{Fe-r} shows the thermal conductivity 
of iron plasma in the density range 
appropriate to the outer envelope of 
a neutron star at $\lg T$[K]= 6.5, 7.0, 7.5 and 8.0. 
Fig.~\ref{all-r} shows the thermal conductivity 
of matter composed of H, He, C, Fe at $T=10^7$ K 
in about the same density range 
(related to the outer envelope of a 
neutron star or to the core of a white dwarf). 
We assume that no impurities are present, 
but we include the contribution of \emph{ee} scattering 
in addition to the \emph{ei} scattering. 
The \emph{ee} scattering contributes to $\nu_\kappa$, i.e, 
lowers the thermal conductivity. 
According to Paper~I,
\beq
     \nu^{ee}  =  {3 \alpha_{\rm f}^2 \, (k_{\rm B} T)^2
          \over 2 \pi^3 \,
       \hbar m_{\rm e}^\ast c^2} 
       \left( {2k_{\rm F}\over k_{\rm TF}}\right)^3
       J(x_{\rm r},y),
\label{ee}
\eeq
where
$y=\sqrt{3} \, T_{\rm pe}/T$,
$T_{\rm pe}=(\hbar/\kB)\,
\sqrt{ { 4 \pi e^2 n_{\rm e} / m_{\rm e}^\ast}}$
is the electron plasma temperature, and
\begin{eqnarray}
J(x_{\rm r},y) &\approx& \left( 1 + {6 \over 5 x_{\rm r}^2} +
     {2 \over 5 x_{\rm r}^4} \right) \,
     \left[ {y^3 \over 3 (1+ 0.07414 \, y)^3} \right.
\nonumber \\
   && \times 
     \ln \left( 1 + {2.81\over y} 
     - {0.81\over y} \, {v_{\rm F}^2\over c^2} \right)
\nonumber\\&&    \left.   
  + {\pi^5 \over 6} \, {y^4 \over (13.91 +y)^4} \right].
\label{Jfit}
\end{eqnarray}
For a plasma of light elements, $Z \la 6$,
the strongest effect of the \emph{ee}
collisions takes place at temperatures comparable 
to degeneracy temperature $T_{\rm F}$
(Lampe \cite{lampe68}; Urpin \& Yakovlev \cite{uy80}),  
as confirmed by Fig.\ \ref{all-r}. 
For Fe matter, \emph{ee} collisions  
are unimportant. 
For H and He plasmas,
their effect is more pronounced at lower $\rho$, where 
the chosen temperature, $T=10^7$ K, is closer to $T_{\rm F}$. 
We conclude that \emph{ee} collisions do not play 
a leading role in thermal transport by degenerate electrons 
but should be taken into account  
in a plasma composed of light elements. 
 
% Fig.  %%%%%%%%%%%%%%%%%%%%%%%% FIGURE %%%%%%%%%%%%%%%%%%%%%%%%%%%%%% 

The strong effect of chemical 
composition (Fig.\ \ref{all-r}) is 
related to 
the $Z$-dependence of the collision frequencies 
$\nu^{ei}_{\sigma,\kappa}$. 
The higher is $Z$, the larger is $\nu^{ei}_{\sigma,\kappa}$,  
and the lower is the conductivity. 
Comparing Figs.\ \ref{Fe-r} and \ref{all-r} we see 
that a temperature variation by a factor of 30
can change the thermal conductivity of iron plasma much less 
than altering the chemical composition from 
H to Fe at fixed $T$. This effect has important consequences 
for the relationship between surface and internal temperatures 
of neutron stars (Paper~I). 
 
% Fig.  %%%%%%%%%%%%%%%%%%%%%%%% FIGURE %%%%%%%%%%%%%%%%%%%%%%%%%%%%%% 
\begin{figure}[ht] 
\begin{center} 
 \leavevmode 
 \epsfysize=95.5mm 
\epsffile[27 10 377 407]{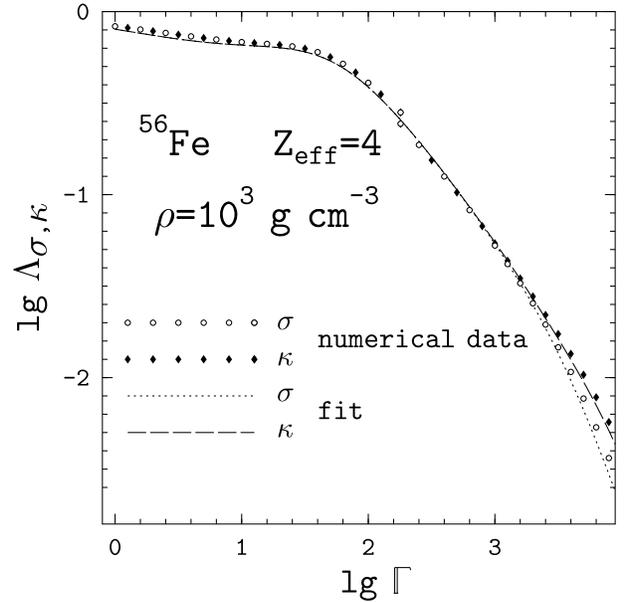} 
\end{center} 
\vspace{-0.4cm} 
\caption[]{ 
Coulomb logarithms versus effective Coulomb 
coupling parameter for a partially ionized iron 
plasma with effective charge $Z_{\rm eff}=4$ 
at $\rho=10^3$ g cm$^{-3}$ in the mean-ion 
approximation. 
} 
\label{L-Fe} 
\end{figure} 

Finally let us consider partially 
ionized matter. We expect that this case can be considered 
in the mean-ion approximation (Sect.\ \ref{sect-numres}). 
Fig.~\ref{L-Fe} shows the dependence of the Coulomb 
logarithms $\Lambda_\sigma$ and $\Lambda_\kappa$ on the 
effective Coulomb plasma parameter 
$\Gamma= Z_{\rm eff}^2 e^2 
(4 \pi n_{\rm i}/3)^{1/3}(\kB T)^{-1}$ for a partially 
ionized Fe matter at $\rho=10^3$ g cm$^{-3}$ and 
$Z_{\rm eff} = 4$. The assumption that $Z_{\rm eff}$ 
is independent of temperature is unrealistic, and we 
adopt it for illustration only. 
Open and filled symbols show numerical values of 
$\Lambda_\sigma$ and $\Lambda_\kappa$, respectively, 
calculated in the mean-ion approximation, while 
dotted and dashed lines are the fit curves. 
Although we did not include the data with $Z \ll A/2$ 
in the fitting, our fit formulae appear to be robust and 
reproduce them quite well. 
 
%%%%%%%%%%%%%%%%%%%%%%%%%%%%%%%%%%%%%%%%%%%%%%%%%%%%%%%%%%%%%%%% 
\section{Conclusions} 
\label{sect-concl} 
We have reconsidered the electrical and thermal 
conductivities of degenerate electrons 
produced by \emph{ei} scattering, 
using the modified structure factors of ions 
as suggested in Paper~II.  
We have analysed the electron transport 
in a wide range of densities, from about $10^2 - 10^4\gcc$ 
to $\sim10^7-10^{10}\gcc$, and temperatures $T\sim10^4-10^9$~K,  
for chemical compositions of astrophysical importance. 
The obtained conductivities differ 
significantly from those 
calculated previously in a wide range of temperatures, 
$T_{\rm m}/5 \la T \la 5T_{\rm m}$, near the melting 
temperature $T_{\rm m}$ of Coulomb crystals. 
Our new approach has reduced 
unrealistically large jumps of the transport 
coefficients at the melting point obtained in the earlier 
works. This, in turn, allowed us to 
develop a unified description of electron conduction 
in liquid and crystal matter and obtain an effective 
potential for the \emph{ei} interaction. 
 
The \emph{ei} scattering, which we studied in this article, 
is known to be the most important mechanism 
of electron relaxation under prevailing physical conditions in 
the envelopes of neutron stars and in the cores of white 
dwarfs. We expect that the improved transport coefficients 
can be used to solve various problems of 
the physics of neutron stars and white dwarfs 
(cooling, evolution of accreting 
stars, nuclear burning of matter, pulsation modes, 
evolution of magnetic fields, etc.). 
 
\begin{acknowledgements} 
We thank F.J.\ Rogers and H.E.\ DeWitt for unpublished tables 
of the static structure factor of ion liquid.  
D.G.Y., A.Y.P.\ and D.A.B.\ are grateful  
for the hospitality of and stimulating 
atmosphere at the N.~Copernicus Astronomical Center in Warsaw. 
This work was supported by Grant Nos.\ RFBR 96--02--16870a, 
DFG--RFBR 96--02--00177G, INTAS 96--0542 and KBN 2 P03D 014 13.  
\end{acknowledgements} 
 
% ==================================================================== 
\appendix 
\addtocounter{section}{1} 
\setcounter{equation}{0} 
\section*{Appendix: analytical fits to Coulomb logarithms} 
Carrying out the integration in \req{L} with 
the effective potential given by \req{Ufit} 
yields the following 
expressions for the Coulomb logarithms, which enter 
\req{nusk} for the effective collision frequencies: 
\beq 
        \Lambda_{\sigma,\kappa}^{\rm fit} = 
               \left[\Lambda_1(s,w) - 
	{v^2\over c^2} \Lambda_2(s,w)\right] G_{\sigma,\kappa}(\eta,\beta). 
\label{Lambdafit} 
\eeq 
Here $s = q_{\rm s}^2/(2 k_{\rm F})^2$, 
$w = w(2k_{\rm F})$ [see Eqs.\ (\ref{qs2})--(\ref{Deta})], 
\begin{eqnarray} 
       2 \Lambda_1(s,w) &=&  \ln{s+1 \over s} + {s\over s+1} 
	  \, (1-\mathrm{e}^{-w})  
\nonumber\\ 
    & - & (1+s w) \, \mathrm{e}^{s w}  
     \left[ {\rm E}_1\,(s w) - {\rm E}_1\,(sw + w) \right], 
\\ 
  2 \Lambda_2(s,w) &=& \!\! {\mathrm{e}^{-w} - 1 + w \over 
	w} - {s^2 \over s+1} 
	   (1-\mathrm{e}^{-w}) - 2 s \ln{s+1 \over s} 
\nonumber\\ 
    & + &    s (2 + s w) \, \mathrm{e}^{s w}  
    \left[ {\rm E}_1\,(s w) - {\rm E}_1\,(s w + w) \right], 
\end{eqnarray} 
and $E_1(x) = \int^{\infty}_x y^{-1}\mathrm{e}^{-y}{\rm\,d}y$ 
is the exponential integral given, for example, 
by the rational-polynomial approximations in  
Abramowitz \& Stegun (\cite{Abramowitz}). 
 
In the particular case of $s \ll 1$ and $s \ll w^{-1}$, 
numerical cancellation  
of large numbers can be avoided with aid of the 
asymptotic expressions 
\begin{eqnarray} 
     \Lambda_1(s\to0,w) &=& \frac12 \left[ 
     E_1(w)+\ln w+\gamma  \right], 
\\ 
     \Lambda_2(s\to0,w) &=& 
     {\mathrm{e}^{-w}-1+w \over 2w}, 
\end{eqnarray} 
where $\gamma=0.5772\ldots$ is the Euler's constant. 
 
In the limiting case of $w \ll 1$, we obtain 
\begin{eqnarray} 
     \Lambda_1(s,w\ll1) &\approx& w \left( {2s+1\over 2s+2} 
     - s\,\ln{s+1\over s} \right), 
\\ 
     \Lambda_2(s,w\ll1) &\approx& 
     w \left( {1-3s-6s^2 \over 4s+4 } + 
     \frac32\,\ln{s+1\over s} \right). 
\end{eqnarray} 

In the opposite case of $ w \gg 1$, 
the familiar expressions for the Debye-like screening 
are recovered: 
\begin{eqnarray} 
     \Lambda_1(s,w \gg 1) &=& \frac12\left( 
           \ln{s+1\over s} - {1\over s+1} \right), 
\\ 
     \Lambda_2(s,w \gg 1) &=& {2s+1\over 2s+2} 
         - s\,\ln{s+1\over s}. 
\end{eqnarray} 

\end{document}